\documentclass[prd,aps,a4paper,twocolumn,eqsecnum,nofootinbib,floatfix]{revtex4}  

\newif\ifusesec
\usesectrue  
   
\usepackage{graphicx} 
\usepackage{amsmath,amsfonts,amssymb}

\newcommand{\beq}{\begin{equation}}
\newcommand{\eeq}{\end{equation}}
\newcommand{\bea}{\begin{eqnarray}}
\newcommand{\eea}{\end{eqnarray}}
\newcommand{\eq}{\eqref}

\newcommand{\Int}{\int\limits}

\newcommand{\g}{{\gamma}}

\newcommand{\pinf}{p_{\infty}}

\def\Z#1{\zeta(#1)}
\def\Q#1{\mathrm{Q}_{#1}}
\def\A#1{{a}_{#1}}
\def\cat#1{{\beta(#1)}}
\def\catd{{\rm K}}
\def\catdd{{{\rm K}^2}}

\begin{document}

\title{
Gravitational dynamics at $O(G^6)$: \\
perturbative gravitational scattering meets experimental mathematics
}

\author{Donato Bini$^{1,2}$, Thibault Damour$^3$, Andrea Geralico$^1$, Stefano Laporta$^{4,5}$, Pierpaolo Mastrolia$^{4,5}$}
  \affiliation{
$^1$Istituto per le Applicazioni del Calcolo ``M. Picone,'' CNR, I-00185 Rome, Italy\\
$^2$INFN, Sezione di Roma Tre, I-00146 Rome, Italy\\
$^3$Institut des Hautes Etudes Scientifiques, 91440 Bures-sur-Yvette, France\\
$^4$Dipartimento di Fisica ed Astronomia, Universit\`a di Padova, Via Marzolo 8, 35131 Padova, Italy\\
$^5$INFN, Sezione di Padova, Via Marzolo 8, 35131 Padova, Italy
}

\date{\today}

\begin{abstract}
A recently introduced approach to the gravitational dynamics of binary systems involves intricate integrals, linked
to nonlocal-in-time interactions arising at the 5-loop level of  classical gravitational scattering. 
We complete the analytical evaluation of classical  gravitational
scattering at the sixth order in Newton's constant, $G$, 
and at the sixth post-Newtonian accuracy. 
We use computing techniques developed for the evaluation of multi-loop Feynman integrals
to obtain our results  in two ways: high-precision arithmetic, yielding
reconstructed analytic expressions, and direct integration {\it via} Harmonic Polylogarithms.
The analytic expression of the tail contribution to the scattering  involve transcendental constants up to
weight four.
\end{abstract}

\maketitle

\section{Introduction}

The detection of the gravitational wave  signals emitted by compact
binary systems~\cite{LIGOScientific:2018mvr} has opened a new
path for investigating the structure of the Universe, and offers a
novel tool for studying the gravitational interaction. The need to
model with high accuracy the gravitational wave signals
emitted during the last orbits of coalescing black-hole binaries
 motivates the development of ever more accurate methods for describing 
 the dynamics and radiation of gravitationally interacting binary systems. 

The present work will demonstrate how progress in the theoretical description of binary
systems can be reached by combining methods developed in General
Relativity (GR) with ideas borrowed from computational methods in Quantum
Field Theory (QFT). Specifically, we show how to complete a state-of-the-art approach to the classical
dynamics of binary systems \cite{Bini:2019nra} by using advanced computing techniques developed for the evaluation of
multi-loop Feynman integrals (see, e.g.,
Refs. \cite{Laporta:2001dd,Laporta:2003jz,Laporta:2017okg,Laporta:1993qw,Laporta:1993ds,Remiddi:1999ew,Argeri:2007up}
for reviews and significant applications) 

The approach of \cite{Bini:2019nra} extracts information from various classical GR observables. In particular, one of the
crucial gauge-invariant observables used in this approach is the classical scattering angle $\chi^{\rm tot}$ 
during a gravitational encounter,
considered as a function of the total center-of-mass energy, $E=\sqrt{s}$, the total angular momentum, $J$, and the symmetric mass
ratio $\nu=\frac{m_1m_2}{(m_1+m_2)^2}$. The approach of \cite{Bini:2019nra} decomposes $\chi^{\rm tot}(E,J ;\nu)$ into three separate contributions: 
\beq \label{chitot0}
\chi^{\rm tot}=\chi^{\rm loc, f}+ \chi^{\rm nonloc, h}+ \chi^{\rm f-h} \ .
\eeq
Here $\chi^{\rm loc, f}$  is the scattering angle that would be induced by the local-in-time piece of the Hamiltonian, 
$H^{\rm loc,f}(t)$, given, at some perturbative accuracy, by a function of the instantaneous state of the binary system.
By contrast, $\chi^{\rm nonloc, h}$, which will be
the focus of the present work, is induced by the nonlocal-in-time piece of the Hamiltonian, $H^{\rm nonloc,h}(t)$ (defined below). 
The nonlocal-in-time scattering angle  $\chi^{\rm nonloc, h}$ is obtained by a classical perturbative method which expands it in
combined powers of the gravitational constant $G$ and of the inverse velocity of light $\frac1c$. At the sixth post-Minkowskian (6PM)
level, i.e., at order $G^6$ (corresponding to 5-loop graphs in the diagrammatic representation of the classical scattering angle
\cite{Damour:2017zjx}), the approach of \cite{Bini:2019nra} gave integral representations of $\chi^{\rm nonloc, h}$
(recalled below)  that were too complicated to be evaluated by the methods ordinarily used in GR. In the present work, we use 
advanced computing techniques developed for the evaluation of multi-loop QFT integrals to derive, for the first time, 
the analytical value of the $O(G^6)$ contribution to $\chi^{\rm nonloc, h}$. In addition, we use the so-acquired
knowledge to correspondingly derive the, hitherto lacking, analytical expression of the $O(G^6)$ contribution to
the last part, $\chi^{\rm f-h}$ (defined below), of the total scattering angle, Eq. \eq{chitot0}.

\section{Classical perturbative expansion of the nonlocal-in-time scattering angle}

The  approach of \cite{Bini:2019nra} (further developed in \cite{Bini:2020wpo,Bini:2020nsb,Bini:2020hmy})
 is based on a novel way of combining results from several theoretical formalisms, developed for studying the
gravitational potential within classical GR: post-Newtonian (PN) expansion, post-Minkowskian (PM) expansion,
multipolar-post-Minkowskian expansion, effective-field-theory,
gravitational self-force approach, and effective one-body method. 
Within this framework, the Hamiltonian of binary systems is decomposed into three types of contributions, 
\beq \label{Htot}
H^{\rm tot}(t)=H^{\rm loc,f}(t)+ H^{\rm nonloc,h}(t)+ H^{\rm f-h}(t)\,.
\eeq
Here, $H^{\rm loc,f}(t)$ is a local-in-time Hamiltonian, expressed, at some given PN accuracy, by an algebraic function of 
the instantaneous values, $q(t), p(t)$, of position and momenta variables. By contrast,  $H^{\rm nonloc,f}(t)$ is a
nonlocal-in-time Hamiltonian, which involves integrals over (at least one) auxiliary time-shifted variable $t'=t+\tau$:
\beq \label{Hnonloch}
H^{\rm nonloc,h}(t) = \frac{G {\cal M}}{c^3} {\rm Pf}_{2 r_{12}^h(t)/c} \int_{- \infty}^{+ \infty}  \frac{ dt'}{|t-t'|} {\cal F}_{\rm GW}^{\rm split}(t,t')+ \cdots
\eeq
Here,  ${\cal M} $ ($= \frac{E}{c^2}$) denotes the total conserved (center-of-mass) mass-energy of the binary system; 
${\rm Pf}_{2 r_{12}^h(t)/c}$ denotes the partie-finie regularization, using the time scale $\Delta t^h= 2 r_{12}^h(t)/c$,
of the logarithmically divergent $t'$ integration at $t'=t$; $ r_{12}^h(t)$ denotes
 the harmonic-coordinate distance between the two bodies; and ${\cal F}_{\rm GW}^{\rm split}(t,t')$ is a
time-split version of the  gravitational-wave energy flux (absorbed and then) emitted  by the system\footnote{We consider
the conservative dynamics of a binary system interacting in a time-symmetric way.}. The ellipsis in Eq. \eq{Hnonloch}
denotes higher-order tail effects, containing higher powers of $ \frac{G {\cal M}}{c^3}$, such as the  
 second-order tail ($\propto \left(\frac{G {\cal M}}{c^3}\right)^2$) analytically derived in \cite{Bini:2020hmy}
 up to the combined 6PM and 5.5PN accuracy. The local Hamiltonian
$H^{\rm loc,f}$ starts at Newtonian level, while $H^{\rm nonloc,h}$
gets contributions from the 4PN order on, and its structure is known up
to 6PN \cite{Bini:2020wpo,Bini:2020nsb,Bini:2020hmy}.
Finally, the last term $H^{\rm f-h}(t)$ is a local-in-time contribution which involves 
the (unsplit) gravitational wave energy flux ${\mathcal F}_{\rm GW}(t)={\cal F}_{\rm GW}^{\rm split}(t,t)$, and 
a  flexibility factor $f(t)= 1+ O\left(\frac1{c^2} \right)$ that is a function of the instantaneous state of the system:
\beq \label{Hfmh}
 H^{\rm f-h}(t)= + 2 \frac{G {\cal M}}{c^3}{\mathcal F}^{\rm GW}(t)  \ln \left( f(t)\right)\,.
\eeq
The latter Hamiltonian contribution is local-in-time, but the determination of the flexibility factor $f(t)$ depends on the
explicit knowledge of the scattering angle induced by the nonlocal-in-time Hamiltonian $H^{\rm nonloc,h}(t) $.

The Hamiltonian decomposition \eq{Htot} yields a corresponding decomposition of the total
scattering angle $\chi^{\rm tot}$, as displayed in Eq. \eq{chitot0}.
At the needed accuracy, the nonlocal contribution $\chi^{\rm nonloc, h}$ can be written as \cite{Bini:2017wfr},
\beq \label{chinonloch}
\chi^{\rm nonloc,h}(E,J,\nu)= \frac{\partial W^{\rm nonloc,h}(E,J,\nu)}{\partial J}\,,
\eeq
where,
\beq \label{Wnonloch}
W^{\rm nonloc,h}(E,J ; \nu) =\int_{- \infty}^{+ \infty}dt\, H^{\rm nonloc,h}(t)\,,
\eeq
is the integrated nonlocal action.
Inserting Eq. \eq{Hnonloch} into Eq. \eq{Wnonloch}, one sees that the knowledge of $\chi^{\rm nonloc,h}(E,J;\nu)$
depends on the evaluation of a (regularized) two-fold integral,
\bea \label{Wnonloch2}
W^{\rm nonloc,h}(E,J;\nu) &=& \frac{G E}{c^5}\times \nonumber\\
&&{\rm Pf}_{2 r_{12}^h(t)/c} \Int_{- \infty}^{+ \infty}  
\Int_{- \infty}^{+ \infty} \frac{dt dt'}{|t-t'|} {\cal F}_{\rm GW}^{\rm split}(t,t') \nonumber\\
&&+ \cdots
\eea
The latter integral is to be evaluated along an hyperbolic-motion solution of the local-in-time Hamiltonian $H^{\rm loc,f}(t)$.

The  method of \cite{Bini:2019nra} requires as crucial input the explicit knowledge 
of the double, PM and PN, perturbative expansion of  $W^{\rm nonloc,h}(E,J; \nu)$, i.e., its combined expansion
in powers of  $G$ (PM expansion), and of $\frac1{c^2}$ (PN expansion).
It is convenient  to express the combined PM$+$PN expansion of 
$W^{\rm nonloc,h}(E,J; \nu)$ in terms  of the dimensionless variables
\beq
\pinf \equiv \sqrt{\g^2-1} \ , \quad {\rm and \quad } j \equiv \frac{c J}{G m_1 m_2}\,,
\eeq
where the effective-one-body specific energy $\g $ is defined as 
\beq
\g = \frac{\mathcal{E}_{\rm eff}}{\mu c^2} \equiv \frac{E^2-m_1^2c^4-m_2^2c^4}{2 m_1 m_2 c^4}\,.
\eeq
As $j \propto \frac{c}{G}$, 
the PM expansion of $W^{\rm nonloc,h}$ is equivalent to an expansion in inverse powers of $j$, and reads
(after setting aside the  second-order tail contribution)
\bea \label{Wexp}
\frac{c \,W^{\rm nonloc,h}(\g,j;\nu)}{2\, G \, m_1 m_2}&=&- \nu \pinf^4\left(\frac{A_0^h(\pinf,\nu) }{3 j^3}+ \frac{A_1^h(\pinf,\nu) }{4 \pinf j^4}\right. \nonumber\\
&&\left. +  \frac{A_2^h(\pinf,\nu) }{5 \pinf^2 j^5} + O\left(\frac1{j^6} \right) \right)\,.
\eea
Using Eq. \eq{chinonloch}, this corresponds to the following PM expansion of the corresponding nonlocal scattering angle
\bea \label{chiexp}
\frac12 \chi^{\rm nonloc,h}(\g,j;\nu)&=&+ \nu \pinf^4\left(\frac{A_0^h(\pinf,\nu) }{ j^4}+ \frac{A_1^h(\pinf,\nu) }{ \pinf j^5}\right. \nonumber\\
&&\left. +  \frac{A_2^h(\pinf,\nu) }{ \pinf^2 j^6} + O\left(\frac1{j^7} \right)\right).
\eea
The dimensionless coefficients $A_m^h(\pinf,\nu)$, $m=0,1,2,\cdots$, then admit a PN expansion, i.e., an expansion in powers of 
$\pinf =O\left(\frac1{c}\right)$, modulo logarithms of $\pinf$, say
\beq 
A_m^h(\pinf,\nu)= \sum_{n\geq0} \left[A_{m n}(\nu) + A^{\ln}_{m n}(\nu) \ln \left(\frac{p_{\infty}}{2} \right) \right] \pinf^n\,.
\eeq
The coefficient $A_{m n}(\nu)$ parametrizes a term of order 
$\frac{\pinf^{4+n-m}}{j^{4+m}} \sim\frac{ G^{4+m} }{c^{8+n}}$ (with $m \geq0$, $n \geq0$)
in the combined PM$+$PN expansion of the nonlocal scattering angle.
The leading-order contribution to the nonlocal dynamics is at the combined 4PM and 4PN level, i.e., $\propto G^4/c^8$ 
\cite{Blanchet:1987wq}. The corresponding nonlocal scattering coefficient, coming from $m=0$ and $n=0$, is  
$A_{0}^h(\pinf,\nu)= \pi\left[-\frac{37}{5}\ln \left(\frac{p_{\infty}}{2} \right)-\frac{63}{4}\right]+ O(\pinf^2)$ \cite{Bini:2017wfr}.
The higher-order logarithmic coefficients $A^{\ln}_{m n}(\nu)$ were analytically determined \cite{Bini:2020wpo,Bini:2020nsb,Bini:2020hmy}
so that we shall henceforth focus on the non-logarithmic coefficients $A_{m n}(\nu)$. 
The classical GR perturbative approach of \cite{Bini:2019nra,Bini:2020wpo,Bini:2020nsb,Bini:2020hmy} yields
 explicit integral expressions for the non-logarithmic coefficient $A_{m n}(\nu)$ with integrands that are  polynomials
in the symmetric mass ratio $\nu$. Writing  $A_{m n}(\nu)= \sum_k A_{m n k} \nu^k$, with $k=0,1,2,\cdots$, this finally yields 
explicit, parameter-free double-integral expressions
for the (numerical) coefficient $A_{m n k}$ of $\nu^k$ in the
polynomial $A_{m n}(\nu)$, say
\beq \label{Amnkint}
A_{m n k} =  \int_{- 1}^{+ 1}   \int_{- 1}^{+ 1}  \frac{dT dT'}{|T-T'|}a_{mnk}(T,T')\,.
\eeq 
The structure of the integrands $a_{mnk}(T,T')$ reads
\bea \label{amnk}
a_{mnk}(T,T')&=&R_0^{m n k}(T,T')\nonumber\\
&+&  R_1^{m n k}(T,T') \left({\rm arctanh}(T)-{\rm arctanh}(T')\right)\nonumber\\
&+&  R_2^{m n k}(T,T') \left({\rm arctanh}(T)-{\rm arctanh}(T')\right)^2\nonumber\\
&+&R_3^{m n k}(T,T') \left({\rm arctan}(T)-{\rm arctan}(T')\right)\,,\nonumber\\
\eea
where the coefficients $ R_N^{m n k}(T,T')$ are  rational functions of
$T$ and $T'$. The integration variables are related via $T= {\rm tanh }\frac{v}{2}$ and $T'= {\rm tanh }\frac{v'}{2}$
to the ``hyperbolic eccentric anomalies'' $v$ and $v'$ that parametrize the original time variables
$t$ and $t'$ via the relativistic generalization \cite{DD1985,Cho:2018upo} of the Keplerian representation
of hyperbolic motion. The latter notably involves a relativistic version of the  hyperbolic
Kepler equation: ${\bar n} (t-t_0)=e_t \sinh v-v+ O(\frac1{c^4})$.

It was possible to analytically compute  the numerical coefficients 
$A_{m n k}$ appearing at the 4PM ($G^4$) and 5PM ($G^5$) levels (i.e.,
for $m=0,1$), up to the 6PN, i.e., $\frac1{c^{12}}$ accuracy. 
By contrast, the integrands of Eq. \eq{Amnkint} become so involved at the
6PM order (corresponding to 5-loop classical scattering diagrams), that the use of 
 standard GR integration methods failed to give the analytical values of the 6PM scattering
coefficients $A_{2 2 0}$, $A_{2 4 0}$, $A_{2 4 1}$, $A_{2 4 2}$. Even the numerical evaluation of the
latter coefficients in \cite{Bini:2020hmy} met with difficulties and only produced 8-digit-accurate results.

The lack of analytical determination of the 6PM coefficients $A_{2 2 0}$, $A_{2 4 0}$, $A_{2 4 1}$, $A_{2 4 2}$
is an imperfection that limits the application of the method of \cite{Bini:2019nra} at the 6PN level. In particular, the combination
\beq \label{DvsA2nk}
D = 
\frac1{\pi}\left(\frac52 A_{221}+\frac{15}{8}A_{200}+A_{242}\right)\,,
\eeq
crucially enters the definition  of the flexibility factor $f(t)$, and thereby the analytical definition of the third contribution, 
$ H^{\rm f-h}(t)$, Eq. \eq{Hfmh}, to the total Hamiltonian. The coefficient $D$, Eq. \eq{DvsA2nk}, is of direct physical
significance for the dynamics of coalescing binary systems because it enters
the elliptic-motion observables (such as periastron precession).

We achieve here the important goal of analytically determining all the 6PM scattering coefficients,  $A_{2 n k}$ (and thereby
also the coefficient $D$,  Eq. \eq{DvsA2nk}),
by applying to the integral representations, Eqs. \eq{Amnkint}, \eq{amnk}, some of the
high-precision numerical techniques and analytic methods that have been developed
for evaluating QFT observables, expressed in terms multi-loop Feynman integrals.

\section{ $A_{2 n k}$ and companion coefficients}

In the following we use the notation of Ref.\cite{Bini:2020hmy} and parametrize the 
 (non-logarithmic) scattering coefficients $A_{2nk}$ in terms of the equivalent set of coefficients denoted
 $d_{nk}$, and related to them via\footnote{For simplicity, we shall not use here the coefficients $c_{nk}$
 that are related to the $d_{nk}$'s  through Eq. (4.14) of Ref. \cite{Bini:2020hmy}.}
\bea \label{Ankvsdnk}
\pi^{-1} A_{200} &=& d_{00}, \nonumber\\
\pi^{-1} A_{220} &=& d_{20}+3 d_{00}, \nonumber\\
\pi^{-1} A_{221} &=& d_{21}-2 d_{00}, \nonumber\\
\pi^{-1} A_{240}  &=& d_{20}+d_{40}+\frac32 d_{00}, \nonumber\\
\pi^{-1} A_{241}  &=& d_{21}-\frac{11}{2}d_{00}+d_{41}-2d_{20}, \nonumber\\
\pi^{-1} A_{242}  &=& d_{42}-2d_{21}+3d_{00}\,.
\eea
The coefficients $d_{00}$ and $d_{21}$ (and therefore $A_{200}$ and $A_{221}$) were computed analytically \cite{Bini:2020hmy},
 \bea \label{d00d21}
d_{00}&=&-\frac{99}{4}-\frac{2079}{8}\, \zeta(3)\,,\nonumber\\
d_{21}&=&\frac{1541}{8}+306\, \zeta(3)\,.
\eea
In addition,  some parts of the integrals giving $d_{20}$, $d_{40}$,
$d_{41}$ and $d_{42}$ could be analytically evaluated, leaving as remaining unknown
coefficients the quantities $Q_{20}$, $Q_{40}$, $Q_{41}$ and $Q_{42}$ related
to $d_{20}$, $d_{40}$, $d_{41}$ and $d_{42}$ (and thereby to $A_{2 2 0}$, $A_{2 4 0}$, $A_{2 4 1}$, and $A_{2 4 2}$)
 via  
\bea
\pi d_{20} &=& \left(\frac{32813}{192}+\frac{66999}{224}\ln(2)\right)\pi\nonumber\\
&&+\frac52 Q_{20}-\frac{25883}{720}-\frac{22333}{56}{\rm K}\,,\nonumber\\
\pi  d_{40} &=& \left(\frac{293499}{512}+\frac{442237}{2688}\ln(2)\right)\pi\nonumber\\
&&+\frac52 Q_{40}-\frac{750674317}{762048}-\frac{442237}{2016}{\rm K}\,,\nonumber\\
\pi  d_{41} &=& \left(-\frac{4431841}{48384}-\frac{28735}{64}\ln(2)\right)\pi\nonumber\\
&&+\frac52 Q_{41}+\frac{703435949}{635040}+\frac{28735}{48}{\rm K} \,,\nonumber\\
\pi  d_{42} &=& \left(\frac{1105777}{6048}-\frac{4497}{32}\ln(2)\right)\pi\nonumber\\
&&+\frac52 Q_{42}+\frac{59610947}{317520}+\frac{1499}{8}{\rm K}\,.
\eea
Here {\rm K} denotes  Catalan's constant, defined as 
 ${\rm K} \equiv \beta(2) \equiv$ $\sum_{n=0}^\infty (-1)^n/(2n+1)^2=$ $0.915965\ldots$ (with $\beta$ being the
 Dirichlet function).
The coefficients $Q_{20}$, $Q_{40}$, $Q_{41}$ and $Q_{42}$ are all
expressed as two-fold integrals of the type indicated 
in Eq. \eq{Amnkint}. 
The explicit forms of the integrands $q_{nk}(T,T')$, for $(nk)=(20),(40)$, and $(41)$,
yielding the corresponding coefficients  $Q_{nk}$ after integration on the  square $(T,T')\in [-1,1]\times [-1,1]$,
 are given in the Supplemental Material of the present paper. See below for  the integrand $q_{42}(T,T')$ of $Q_{42}$\footnote{
The original form of  $q_{42}(T,T')$ was presented in the Ancillary File of \cite{Bini:2020hmy}.}.

In their original forms, the integrands $q_{nk}(T,T')$ (including the factor $\frac1{ |T-T'| }$ pulled out  in Eq. \eq{Amnkint})
are   finite but discontinuous across the first diagonal $T=T'$ of the square. One, however, gets a continuous, and simpler, integrand
by working (as it is allowed) with a symmetrized integrand 
$q_{nk}^{\rm sym}(T,T')=\frac12 \left( q_{nk}(T,T')+ q_{nk}(T',T) \right)$.
It is then found that $q_{nk}^{\rm sym}(T,T')$ has a four-fold symmetry. Namely, it is symmetric under reflection through the
two diagonals of the square $(T,T')\in [-1,1]\times [-1,1]$. In other words, $q_{nm}^{\rm sym}(T,T')=+q_{nm}^{\rm sym}(T',T)
= q_{nm}^{\rm sym}(-T,-T')=q_{nm}^{\rm sym}(-T',-T)$. In addition,  $q_{nm}^{\rm sym}(T,T')$ is continuous (though not differentiable)
across the first diagonal $T=T'$. The value of the integral $Q_{nk}$ can then be obtained from  integrating $q_{nk}^{\rm sym}(T,T')$
on the triangle  $-1<T<1, -1<T'<T$ (or even on the subtriangle obtained by modding out the symmetry under the second diagonal).

For concreteness, let us discuss in detail the structure of the  integrand $q_{42}^{\rm sym}(T,T')$. We have
\bea
&& |T-T'| q_{42}^{\rm sym}(T,T') =r_{42}(T,T') \nonumber\\
&& + {\bar c}_1(T,T') \left[ \frac{ A-A' }{ T-T' }- \frac12 \frac{dA}{dT}- \frac12 \frac{dA'}{dT'} \right] \nonumber\\
&& + {\bar c}_2(T,T') \left[ \left(\frac{ A-A' }{ T-T' }\right)^2- \frac12 \left(\frac{dA}{dT}\right)^2- \frac12 \left(\frac{dA'}{dT'}\right)^2 \right]\,, \nonumber\\
\eea
where $r_{42}(T,T') $, ${\bar c}_1(T,T')$ and ${\bar c}_2(T,T')$ are rational functions, and where we used the shorthand
notation $A \equiv {\rm arctanh}(T)$, and $A' \equiv {\rm
  arctanh}(T')$. 
The integral corresponding to $r_{42}(T,T')$ can be explicitly performed, namely
\bea \label{Q42r}
Q^r_{42}&=& \Int_{- 1}^{+ 1}   \Int_{- 1}^{+ 1}  \frac{dT dT'}{|T-T'|} r_{42}(T,T')  \nonumber\\
&=& - \frac{3463}{20} {\rm K} +\frac{33256213}{340200}\nonumber\\
&& -\pi \left(\frac{8306153213}{7257600}+\frac{19776073}{15120}\ln(2)\right)\,,
\eea
while the rational coefficients ${\bar c}_1(T,T')$ and ${\bar c}_2(T,T')$ can be factorized as
\beq
{\bar c}_1(T,T')= \frac{-16 (1-T^2)^2 (1-T'^2)^2(1-T T')}{945 (1 + T^2)^9 (1 + T'^2)^9(1+ T T')^3} P_1(T,T'),
\eeq
\beq
{\bar c}_2(T,T')= \frac{-16 (1-T^2)^3 (1-T'^2)^3}{315 (1 + T^2)^8 (1 + T'^2)^8( 1+T T')^3} P_2(T,T'),
\eeq
where $P_1(T,T')$ and $P_2(T,T')$ are (symmetric) polynomials in $T$ and $T'$ which are given in  the Supplemental Material.
The total degree in $T$ and $T'$ of $P_1(T,T')$ is 32, while that of $P_2(T,T')$ is 28.
Note that the integrands corresponding to ${\bar c}_1(T,T')$ and ${\bar c}_2(T,T')$ vanish proportionally
to $\frac{(T-T')^2}{|T-T'|}= |T-T'|$ across the first diagonal.

\section{Analytic evaluation of the scattering integrals $Q_{nk}$}
\label{sec:analyticeval}

To determine the analytic expressions of the scattering integrals $Q_{nk}$ (equivalent to the $A_{nk}$'s),
we adopt a two-step strategy:
\begin{enumerate}
\item Experimental Mathematics and Analytic Recognition; 
\item Analytic Integration and Harmonic Polylogarithms.
\end{enumerate}
Such a strategy is often used in the realm of multi-loop Feynman
calculus, when a direct analytic integration seems prohibitive, See
{\it e.g.} Refs. 
\cite{Laporta:2017okg,Laporta:2001dd,Laporta:2003xa,Laporta:1993qw,Laporta:1993ds,Remiddi:1999ew,Argeri:2007up,Argeri:2014qva,DiVita:2019lpl,DiVita:2018nnh,Frellesvig:2019kgj,Laporta:2019fmy,Laporta:2020fog}. 
Previous uses of experimental mathematics and high-precision arithmetics within studies of binary systems
include Refs. \cite{Shah:2013uya,Johnson-McDaniel:2015vva,Foffa:2016rgu}. We note in particular that 
one of  the integrals (in momentum space) contributing to the 4PN-static term of the two-body potential, used in
\cite{Foffa:2016rgu} and originally obtained by analytic recognition \cite{Lee:2015eva},
was later analytically confirmed by direct integration (in position space) \cite{Damour:2017ced}.

In the current work, in order to perform the first step, we started by numerically computing the definite integrals
of $q_{nk}(T,T')$ on the triangle $-1<T<1, -1<T'<T$ 
to a very high precision (with a few hundreds of digits), using a double-exponential change of variables \cite{DoubleExponential}.
Indeed, the latter method is well-adapted to our integrals which are mildly singular on the boundaries of the triangle  $-1<T<1, -1<T'<T$.
The 200-digit  accuracy level that we used was amply sufficient for reconstructing the analytic expressions of 
the $Q_{nk}$'s by using the PSLQ algorithm \cite{PSLQ} and a basis of  transcendental constants indicated
both by the structure of the integrands, and the analytical results \eq{d00d21}, \eq{Q42r}.
Our numerical results are given in Table \ref{table_Qij},  while Table \ref{fits} gives the reconstructed analytic
expressions of the $Q_{nk}$'s, and the related $d_{nk}$'s.
%
%
\begin{table*}
\caption{\label{table_Qij} Numerical values of the $Q_{nk}$ integrals
  with 200-digit accuracy.
}
\begin{ruledtabular}
\begin{tabular}{|l|l|}
$Q_{20}$ &$524.7672921802125843427359557031017584761419995573690119377287112384988398300977120939070371581$\\
	 &$96060831706238995205677052067946783744966475134730111010455883184170170829347212071124106113165$\\
	 &$8613485679$\\
\hline
$Q_{40}$ &$544.4939915701706772258458158548215701355843583332648304959367083415682948158610574285653029862$\\
         &$87084252115923423339364981524722633807905033769432119691717874743144282677041484694939992691447$\\
	 &$2804761699$\\
\hline
$Q_{41}$ &$-1029.52887537403849684626420906288951311349891044967686745420133893415513339408657109916000809$\\
	 &$60027000683311179242081514484014334501266712433925887538266005603952131007506207305140646213006$\\
	 &$5024513617$\\
\hline
$Q_{42}$ &$-802.885057050786642755886295069034459970736865058430654964178895902426423211047940727300850918$\\
	 &$74267871623078435110513965444766798525251182468350940531767163197645060875802781537593191860287$\\
	 &$8433814664$\\
\end{tabular}
\end{ruledtabular}
\end{table*}

%
%
\begin{table*}
\caption{\label{fits} PSLQ reconstruction of the various integrals.
}
\begin{ruledtabular}
\begin{tabular}{|l|l|}
$Q_{20}$ & $ \frac{25883}{1800} + \frac{22333}{140} {\rm K} - \frac{625463}{3360}\pi   - 
     \frac{361911}{560}\pi \ln  2    + \frac{99837}{160}\pi \zeta(3) $\\
$d_{20}$ & $ -\frac{32981}{112} - \frac{9216}{7}\ln 2    +  \frac{99837}{64} \zeta(3)$ \\
\hline
$Q_{40}$ & $\frac{750674317}{1905120} + \frac{442237}{5040} {\rm K}  - \frac{571787}{103680}\pi  
   \frac{7207043}{6720} \pi \ln 2   - \frac{190489}{320} \pi \zeta(3)$\\
$d_{40}$ & $\frac{725051}{1296} + \frac{19920}{7} \ln 2   -  \frac{190489}{128} \zeta(3)$ \\
\hline
$Q_{41}$ & $-\frac{703435949}{1587600} - \frac{5747}{24} {\rm K}  +  
   \frac{1154149}{17280} \pi +  \frac{1897771}{3360} \pi \ln 2   - 
      \frac{306219}{640}  \pi \zeta(3)$\\
$d_{41}$ & $\frac{607867}{8064} + \frac{20224}{21} \ln 2  -  \frac{306219}{256} \zeta(3)$ \\
\hline
$Q_{42}$ & $-\frac{59610947}{793800} - \frac{1499}{20} {\rm K}  - \frac{402163}{2520} \pi +
 \frac{4497}{80}\pi \ln 2   - \frac{11871}{160}\pi \zeta (3)$\\
$d_{42}$ & $-\frac{186743}{864}-\frac{11871}{64}\zeta(3)$ \\
\hline
\end{tabular}
\end{ruledtabular}
\end{table*}

We have {\it a posteriori} checked  that the so-reconstructed exact values of the $d_{nk}$'s
agree (within our estimated error $\pm 1 \times 10^{-8}$) with the values given in Table VI of \cite{Bini:2020hmy}.

Having the semi-analytic expressions given in Table \ref{fits} in
hands, we proceed to the (purely analytical) second step of our strategy. 

We first perform the integration over $T'$, beginning with an integration by-parts
 of the terms that contain, in the denominator, polynomials in $T'$ with integer exponents bigger than one.
We are then left with 
integrals containing powers of ${\rm arctan}(T')$, or ${\rm arctanh}(T')$, in the numerator, and powers of $(T'\pm 1)$,
$(T'\pm i)$, $(T'-T)$ and $(T'+1/T),$ in the denominator.
These integrals are carried out by differentiating with respect
to $T$, repeatedly if needed, until the integration in $T'$ is straightforward.
Thereby, the original integral is obtained as a repeated quadrature in
$T$, whose first layer reads as,
\bea
f(T) &=& 
\int_{-1}^T dT' g(T, T') \nonumber \\
&=& f(T_0) +
\int_{T_0}^T dT 
\frac{\partial}{\partial T} \int_{-1}^{T} dT' g(T,T') \nonumber \\
& = &  f(T_0) +
\int_{T_0}^T dT 
\bigg(g(T,T) + \int_{-1}^T dT' 
\frac{\partial g(T,T')}{\partial T} 
\bigg) \,.
\nonumber
\eea

The integrand $f(T)$ obtained after the integration over $T'$ contains Nielsen 
polylogarithms  \cite{Remiddi:1970aa} (up to weight 3).
For convenience, we fold the integral over the interval  $ T \in {[0,1]}$:
$\int_{-1}^1dT f(T)  = \int_{0}^{1}dT \left[ f(T)+f(-T) \right]  $.

The final integration over $ T \in {[0,1]}$ is performed in three steps.
First, we integrate  by-parts the factors $(T \pm i)^{-n}$ or
$(T \pm 1)^{-n}$ with $n>1$, until $n$ is reduced to $1$. 
Second, we map the resulting integrals containing $T^{-1}$ and $(T \pm 1)^{-1}$
(but not $(T \pm i)^{-1}$) to Harmonic Polylogarithms
(HPLs) \cite{Remiddi:1999ew}.
The HPLs are defined as recursive integrals,
\beq
H_{i_1 i_2 {\ldots} i_n}(x) = 
\int_0^x     {dt_1}\; 
f_{i_1}(t_1)
H_{i_2 {\ldots} i_n}(x)  
\ ,
\eeq
with $f_{\pm 1}(x)=(1\mp x)^{-1}$, $f_0(x)=1/x$, and 
$H_{\pm 1}(x)=\ln(1\mp x)$,
$H_0(x) \equiv \ln (x)$. For a given HPL, $H_{i_1 i_2 {\ldots} i_n}(x)$,
the number $n$ of indices is called its {\it weight}, and  corresponds to the number of 
iterations appearing in its nested integral representation. $H$-functions obey integration-by-parts relations and
shuffle algebra relations which can be used to identify,
weight-by-weight, a minimal (albeit not unique) subset of them to be considered as independent. 
For instance, at weights $w =2, 3$, and $4$ the minimal subsets are
formed by 3, 8, and 18 elements, respectively (see \cite{Maitre:2005uu}
for a {\sc Mathematica} implementation).

Third, we consider the integrals containing $(T \pm i)^{-1}$:
these cannot be directly cast in HPL format.
Therefore, we modify the integrands by a suitable insertion of a parameter $x$, to
be later {\it eliminated}, in order to obtain the original integral back. 
The integral, now function of $x$, will be subsequently reconstructed  by repeated differentiations with respect to $x$ and quadratures (as
explained earlier, in the case of the $T'$ integration). 
Let us show an example of this technique: all the $Q_{nk}$ contain the same 
combination of integrals with $w=4$,
\beq\label{jex1}
J=\int_0^1 dT \frac{16 \,{\mathrm{arctanh}}^3(T) -3 \,{\mathrm{Li}}_3\left[
-\left(\frac{1-T}{1+T}\right)^2\right]}{1 + T^2} \ .
\eeq
Notice that ${\rm arctanh}(T) = -(1/2) {\rm ln}((1-T)/(1+T))$.
We modify the integral \eq{jex1}, to let it acquire a dependence on
the variable $x$, {i.e.} $J \to J(x)$, in the following way:
\bea\label{jex2}
&& J(x)\equiv\int_0^1 dT \, (1-x^2) \times \nonumber \\
&& 
\times  \frac{16\, {\mathrm{arctanh}}^3 (T) -3 \,{\mathrm{Li}}_3\left[
\left(\frac{(1-T)(1-x)}{(1+T)(1+x)}\right)^2\right]}{2i (T+x)(T+1/x)}.
\eea
Then, the original integral is recovered at $x=i$, that is $J=J(i)$.
By differentiating and reintegrating over $x$, $J(x)$ can be
conveniently written in terms of HPL's at weight $w=4$, as, 
\bea\label{jexhpl}
&& i\; J(x)=
\nonumber\\
&& \frac{23}{240}\pi^4 -  21 \ln 2 \,\zeta(3) + \pi^2\ln^2 2 - \ln^4 2
- 24 a_4 
\nonumber\\
&&
+\frac{1}{2} \pi^2 H_{0,-1}(x)
+\frac{1}{2} \pi^2 H_{0,1}(x)
-\frac{3}{2} \pi^2 H_{-1, -1}(x)
\nonumber\\
&&
-\frac{3}{2} \pi^2 H_{-1, 1}(x)
-\frac{3}{2} \pi^2 H_{1, -1}(x)
-\frac{3}{2} \pi^2 H_{1, 1}(x) 
\nonumber\\
&&
- 12 H_{0,-1, -1, -1}(x)
+  6 H_{0,-1, -1, 0}(x)
- 12 H_{0,-1, 1, -1}(x)
\nonumber\\
&&
+  6 H_{0,-1, 1, 0}(x)
- 12 H_{0,1,-1, -1}(x)
+  6 H_{0,1, -1, 0}(x)
\nonumber\\
&&
- 12 H_{0,1, 1, -1}(x)
+  6 H_{0,1, 1, 0}(x)
-  6 H_{-1, -1, -1, 0}(x)
\nonumber\\
&&
-  6 H_{-1, -1, 1, 0}(x)
-  6 H_{-1, 1, -1, 0}(x)
-  6 H_{-1, 1, 1, 0}(x) 
\nonumber\\
&&
-  6 H_{1, -1, -1, 0}(x)
-  6 H_{1, -1, 1, 0}(x)
-  6 H_{1, 1, -1, 0}(x)
\nonumber\\
&&
-  6 H_{1, 1, 1, 0}(x)
+ 12 H_{0,-1, -1}(x) \ln 2
+ 12 H_{0,-1, 1}(x) \ln 2
\nonumber\\
&&
+ 12 H_{0,1, -1}(x) \ln 2
+ 12 H_{0,1, 1}(x) \ln 2 
\nonumber\\
&&
+ \frac{21}{2} H_{-1}(x) \zeta(3)
- \frac{3}{2} H_{0}(x) \zeta(3)
+ \frac{21}{2} H_{1}(x) \zeta(3) \ .
\eea
Using $a_4= {\rm Li}_4(1/2)$, and the values of the HPLs at $x=i$, listed in Table \ref{HPLvalues} (see the Appendix), one finds 
the following value for \eq{jex1}:
\beq
J(i) = -\frac{1}{2}\pi^2\; {\rm{K}} + \frac{9}{2} \pi
\zeta(3)\ = J .
\eeq
Using a similar strategy, the expressions of all the coefficients $Q_{nk}(x)$ can be obtained
analytically. The size of the occurring intermediate expressions, similar to \eq{jexhpl}, is too large to be presented here.
Anyway, the crucial results concern the final analytic expressions for the so-obtained 
$Q_{nk} \equiv Q_{nk}(x=i)$. They are found to be drastically simpler than the intermediate results, and,
as expected, to be in perfect agreement with the semi-analytical expressions discussed earlier, and given in Table \ref{fits}.

\section{Scattering angle and periastron precession at 6PM, $O(G^6)$}

The 6PM-accurate  ($O(G^6)$) scattering coefficient $A_2^h(\pinf,\nu)$ associated with
 the integrated nonlocal action $W^{\rm nonloc,h}$, Eq. \eq{Wnonloch}, 
 when  PN-expanded in powers of $\pinf$, reads,
\bea \label{A2PNexp}
A_2^h(\pinf;\nu)&=&A_2^{\rm tail, h, N}+ A_2^{\rm tail, h, 1PN}+ A_2^{{\rm tail}^2, \rm h, 1.5PN}\nonumber\\
&& + A_2^{\rm tail, h, 2PN} + O(\pinf^5)\,.
\eea
The values of the first, $A_2^{\rm tail, h, N}$, and third, $A_2^{{\rm tail}^2, \rm h, 1.5PN}$, contributions (respectively contributing to the 4PN and 5.5PN orders) were obtained in Ref. \cite{Bini:2020hmy}.
New with the present work is the complete  analytical determination
of the two other  contributions to Eq. \eq{A2PNexp}, namely,
 $A_2^{\rm tail, h, 1PN}$, and $A_2^{\rm tail, h, 2PN}$. The latter contributions are both at the 6PM ($O(G^6)$) order, and they
  respectively belong to the 5PN ($O(c^{-10})$) and  6PN ($O(c^{-12})$) levels. Recalling also the 4PN contribution to $A_2^h(\pinf;\nu)$,
  we have now the complete, 6PN-accurate analytical results for $A_2^{\rm tail, h}$:
\begin{widetext}
\begin{eqnarray}
A_2^{\rm tail, h, N}&=&\pi \left[
-\frac{2079}{8}\zeta(3) -\frac{99}{4}
 -122 \ln \left(\frac{p_{\infty}}{2} \right) 
\right]\,,\nonumber\\
A_2^{\rm tail, h, 1PN}&=&\pi 
\left[\left(-\frac{13831}{56}+\frac{811}{2}\nu\right)\ln\left(\frac{p_{\infty}}{2} \right)-\frac{41297}{112}+\frac{1937}{8}\nu
+\left(\frac{49941}{64} +\frac{3303}{4}\nu \right)\zeta(3)-\frac{9216}{7}\ln(2)
\right]p_{\infty}^2
\,,\nonumber\\
A_2^{\rm tail, h,  2PN}&=&\pi 
\left[\left(\frac{75595}{168}\nu+\frac{64579}{1008}-785\nu^2\right)\ln\left(\frac{p_{\infty}}{2} \right)+\frac{1033549}{4536}+\frac{8008171}{8064}\nu-\frac{583751}{864}\nu^2\right. \nonumber\\
&& -\left(\frac{40711}{128}+\frac{660675}{256}\nu -\frac{100935}{64}\nu^2 \right)\zeta(3)  \nonumber\\
&&\left.+\left(\frac{10704}{7}+\frac{75520}{21}\nu\right)\ln(2)
 \right]p_{\infty}^4
\,.
\end{eqnarray}
\end{widetext}

As a consequence of our results, we can also now compute the analytical value of  the (minimal value of the) flexibility coefficient,
$f(t)$, and thereby the effect of  $H^{\rm f-h}(t)$, Eq. \eq{Hfmh}, on the near-zone gravitational physics, such as
periastron precession. They both depend on the crucial combination $D$, Eq. \eq{DvsA2nk}.
Though all the building blocks entering  $D$ (which can equivalently be written as
$D=\frac12 d_{21}+d_{42}-\frac18 d_{00}$) contain $\zeta(3)$, it is remarkably found that $D$
turns out to be equal to the rational number
\beq
D= -\frac{12607}{108}\,,
\eeq
which is compatible with the previous numerical estimate
$D^{\rm num}= -116.73148147(1)$.

The value of $D$ then determines the minimal value of the flexibility coefficient
$D_3^{\rm min}$ (see Eq. (7.28) in \cite{Bini:2020hmy}), namely 
\bea 
D_3^{\rm min}  = -\frac{68108}{945}\nu \,,
\eea
as well as the $f$-related, 6PN-level contribution to the periastron precession (see Eq. (8.30) in \cite{Bini:2020hmy}):
\begin{eqnarray}
K^{\rm f-h,  circ, min}(j)&=& + \frac{68108}{945}\frac{\nu^3}{ j^{12}} \,.
\end{eqnarray}

\section{Conclusions}

 By using advanced computing techniques developed for the evaluation
 of  multi-loop Feynman integrals, we have completed the analytical  knowledge of classical  gravitational scattering 
 (and periastron precession) at the sixth order in $G$, and at the sixth post-Newtonian  accuracy. 
 We think that the present work exemplifies a new type of synergy between classical GR and QFT techniques that
 can be developed in many directions, and can significantly help to improve the theoretical description
 of gravitationally interacting binary systems.

\section*{Acknowledgments}
DB and TD thank Massimo Bernaschi for collaboration at an early stage of this project, and for informative discussions on numerical integration. 
DB and PM acknowledge the hospitality, and the highly stimulating environment, of the Institut des Hautes Etudes Scientifiques. 
 SL thanks CloudVeneto for the use of computing and storage facilities. DB and AG thank MaplesoftTM for providing a complimentary licence of Maple 2020.

\section*{Appendix}

We collect in Table \ref{HPLvalues}, the values of the independent bases of HPLs,  $H_{i_1 i_2
    {\ldots} i_n}(x)$ at the point $x=i$, up to weight $w=4$, required
  for the analytic evaluations described in Sec. \ref{sec:analyticeval}.

%
%
\begin{table*}
\caption{\label{HPLvalues} Independent sets of HPLs, at the point $x=i$, up to weight four.}
\begin{ruledtabular}
\begin{tabular}{|l|l|}
$H_{-1}(i)   $&$  \frac{\ln 2}{2} +i \frac{\pi}{4} $\\
$H_{0}(i)    $&$                   i \frac{\pi}{2} $\\
$H_{1}(i)    $&$ -\frac{\ln 2}{2} +i \frac{\pi}{4} $\\
\hline
$H_{0,-1}(i)  $&$    \frac{\pi^2}{48} +i \catd $\\
$H_{0,1}(i)   $&$  - \frac{\pi^2}{48} +i \catd $\\ 
$H_{-1, -1}(i)$&$   - \frac{\pi^2}{32} + \frac{\ln^2 2}{8} +\frac{1}{8} i\pi \ln 2$\\ 
$H_{-1, 1}(i) $&$   - \frac{\pi^2}{32} - \frac{\ln^2 2}{8} -\frac{3}{8} i\pi \ln 2  +i \catd $\\ 
$H_{1, -1}(i) $&$   - \frac{\pi^2}{32} - \frac{\ln^2 2}{8} +\frac{3}{8} i\pi \ln 2  -i \catd $\\
$H_{1, 1}(i)  $&$   - \frac{\pi^2}{32} + \frac{\ln^2 2}{8} -\frac{1}{8} i\pi \ln 2$\\
\hline
$H_{0,-1, -1}(i) $&$ \frac{29}{64}\Z3  -\frac{1}{4} \catd \pi - i \Q3 $\\
$H_{0,-1, 1}(i)  $&$ \frac{27}{64}\Z3  -\frac{1}{4} \catd \pi + i \frac{\pi^3}{32} - 3 i \Q3 - 2 i \catd \ln 2 $\\ 
$H_{0,1, -1}(i)  $&$ \frac{27}{64}\Z3  -\frac{1}{4} \catd \pi - i \frac{\pi^3}{32} + 3 i \Q3 + 2 i \catd \ln 2 $\\
$H_{0,1, 1}(i)   $&$ \frac{29}{64}\Z3  -\frac{1}{4} \catd \pi + i \Q3 $\\ 
\hline
$H_{0,-1, -1, -1}(i) $&$  \frac{61}{15360}\pi^4 
   - \frac{35}{128} \Z3 \ln 2  
   +  \frac{5}{384} \pi^2 \ln^2 2 
   -  \frac{5}{384} \ln^4 2 
   - \frac{5}{16} \A4
   + \frac{\pi \Q3}{4} 
   + i \Q4
    $\\   
$H_{0,-1, -1, 0}(i) $&$  -\frac{\pi^4}{4608} + \frac{\catdd}{2} +
\frac{\pi \Q3}{2} + \frac{29}{128} i \pi \Z3
  - \frac{7}{48} i \catd \pi^2 $\\
$H_{0,-1, 1, -1}(i) $&$
  - \frac{97}{9216}\pi^4
   + \frac{91}{128} \Z3 \ln 2 
   - \frac{13}{384} \pi^2 \ln^2 2 
   + \frac{13}{16} \A4 
   - \frac{13}{384}\ln^4 2
   + \frac{3}{4} \pi \Q3 
+   \frac{\catd^2}{2} 
   - \frac{7}{8} i \pi \Z3 
   + \frac{1}{16} i \pi^3 \ln 2 
   - 2 i \catd \ln^2 2 
   $\\&$
- \frac{5}{16} i \catd \pi^2 
  + 5 i \cat4 
   -  6 i \Q3 \ln 2 
   - 9 i \Q4 
   $\\
$H_{0,-1, 1, 0}(i) $&$ 
  - \frac{71}{4608} \pi^4 
  + \frac{3}{2}\pi \Q3
  + \catd \pi \ln 2
  +\frac{\catdd}{2} 
   + \frac{27}{128} i \pi \Z3
  + \frac{1}{8} i \pi^3 \ln 2
  - \frac{5}{48} i \catd \pi^2 
  - 3 i \cat4 
  $\\
   
$H_{0,1, -1, -1}(i) $&$ 
     \frac{169}{9216}\pi^4 
    -  \frac{77}{128} \Z3 \ln 2 
    + \frac{9}{128} \pi^2 \ln^2 2 
    -  \frac{27}{16} \A4
    - \frac{9}{128} \ln^4 2
   - \frac{3}{4} \pi \Q3 
   -  \frac{1}{2} \catd \pi \ln 2
    - \frac{21}{128} i \pi \Z3
   - \frac{1}{32} i \pi^3 \ln 2
   $\\&$
   + i \catd \ln^2 2 
   + i \cat4 
   + 2 i \Q3 \ln 2 
   + i \Q4 
   $\\
$H_{0,1, -1, 0}(i) $&$ 
    \frac{73}{4608}\pi^4 
  +\frac{\catdd}{2}
  - \catd \pi \ln 2
  - \frac{3}{2}\pi \Q3 
   + \frac{27}{128} i \pi \Z3
   -  \frac{1}{8} i \pi^3 \ln 2
  - \frac{7}{48} i \catd \pi^2 
  + 3 i \cat4 
   $\\   
$H_{0,1, 1, -1}(i)  $&$
  \frac{61}{9216}\pi^4 
 + \frac{21}{128} \Z3 \ln 2 
 + \frac{13}{384} \pi^2 \ln^2 2 
 - \frac{13}{384} \ln^4 2
 - \frac{13}{16} \A4
 - \frac{1}{2} \catd \pi \ln 2
 + \frac{\catdd}{2} 
 - \frac{1}{4}\pi \Q3 
 + \frac{133}{128} i \pi \Z3 
 - \frac{1}{32} i \pi^3 \ln 2 
   $\\&$
 + i \catd \ln^2 2 
 + \frac{3}{16} i \catd \pi^2
 - 5 i \cat4 
 + 4 i \Q3 \ln 2 
 + 7 i \Q4
   $\\   
$H_{0,1,1, 0}(i) $&$
  - \frac{\pi^4}{4608} 
  + \frac{\catdd}{2} 
  - \frac{1}{2} \pi \Q3
  - \frac{5}{48} i \catd \pi^2 
  + \frac{29}{128} i \pi \Z3
   $\\   
$H_{-1, -1, -1, 0}(i)  $&$  
    - \frac{31}{15360} \pi^4
   + \frac{1}{2} \Z3 \ln 2 
   - \frac{1}{32} \pi^2 \ln^2 2
   + \frac{5}{384} \ln^4 2
   + \frac{5}{16} \A4 
   + \frac{29}{256} i \pi \Z3 
    + \frac{1}{96} i \pi \ln^3 2 
    - \frac{1}{96} i \pi^3 \ln 2 
    - \frac{1}{32} i \catd \pi^2
   $\\&$
   - \frac{1}{8} i \catd \ln^2 2
   - \frac{1}{2} i \Q3 \ln 2 
    - i \Q4 
   $\\   
$H_{-1, -1, 1, 0}(i)  $&$ 
   - \frac{115}{9216}\pi^4 
   + \frac{13}{16} \Z3 \ln 2 
   - \frac{1}{12} \pi^2 \ln^2 2
   + \frac{9}{128} \ln^4 2 
   + \frac{27}{16} \A4
   + \frac{1}{4} \catd \pi \ln 2
   + \frac{1}{2}\pi \Q3 
   + \frac{27}{256} i \pi \Z3
   - \frac{1}{96} i \pi \ln^3 2
   $\\&$
   + \frac{1}{24} i \pi^3 \ln 2
   - \frac{1}{32} i \catd \pi^2
   - \frac{1}{8} i \catd \ln^2 2
   - i \cat4 
   - \frac{1}{2} i \Q3 \ln 2
   - i \Q4 
   $\\   
$H_{-1, 1, -1, 0}(i) $&$ 
    \frac{91}{9216} \pi^4
  - \frac{1}{2} \Z3 \ln 2 
  + \frac{17}{96} \pi^2 \ln^2 2
  - \frac{13}{384} \ln^4 2
  - \frac{13}{16} \A4 
  - \pi \Q3
  + \frac{\catdd}{2}
  - \frac{3}{4} \catd \pi \ln 2
  + \frac{335}{256} i \pi \Z3
  - \frac{1}{96} i \pi \ln^3 2
   $\\&$
  - \frac{3}{64} i \pi^3 \ln 2 
  + \frac{13}{96} i \catd \pi^2
  + \frac{9}{8} i \catd \ln^2 2
  - 5 i \cat4
  + \frac{9}{2} i \Q3 \ln 2
  + 9 i \Q4
   $\\   
$H_{-1, 1, 1, 0}(i) $&$
  - \frac{79}{9216} \pi^4 
  + \frac{1}{16} \Z3 \ln 2 
  - \frac{7}{48} \pi^2 \ln^2 2
  + \frac{13}{384} \ln^4 2 
  + \frac{13}{16} \A4 
  + \frac{\catdd}{2}
  + \frac{1}{2} \pi \Q3
  + \frac{1}{2} \catd \pi \ln 2
  - \frac{195}{256} i \pi \Z3
  + \frac{1}{96} i \pi \ln^3 2
   $\\&$
  + \frac{3}{64} i \pi^3 \ln 2
  - \frac{31}{96} i \catd \pi^2
  - \frac{7}{8} i \catd \ln^2 2
  + 5 i \cat4
  - \frac{7}{2} i \Q3 \ln 2
  - 7 i \Q4
   $\\   
$H_{1, -1, -1, 0}(i) $&$ 
   \frac{55}{9216} \pi^4
 - \frac{7}{96} \pi^2 \ln^2 2
 - \frac{1}{16} \Z3 \ln 2 
 - \frac{13}{384} \ln^4 2
 - \frac{13}{16} \A4
 + \frac{1}{2} \catd \pi \ln 2
 + \frac{1}{2}\pi \Q3 
 - \frac{\catdd}{2}
 - \frac{279}{256} i \pi \Z3
 - \frac{1}{96} i \pi \ln^3 2
   $\\&$
 - \frac{7}{96} i \catd \pi^2
 - \frac{7}{8} i \catd \ln^2 2
 + 5 i \cat4
 - \frac{7}{2} i \Q3 \ln 2
 - 7 i \Q4 
   $\\   
$H_{1, -1, 1, 0}(i) $&$
   \frac{29}{9216} \pi^4
 + \frac{1}{2} \Z3 \ln 2 
 + \frac{5}{48} \pi^2 \ln^2 2
 + \frac{13}{384} \ln^4 2
 + \frac{13}{16} \A4 
 - \frac{3}{4} \catd \pi \ln 2
 -\frac{\catdd}{2}
 - \pi \Q3
 + \frac{167}{256} i \pi \Z3
 + \frac{1}{96} i \pi \ln^3 2
   $\\&$
 - \frac{1}{32} i \pi^3 \ln 2
 + \frac{37}{96} i \catd \pi^2
 + \frac{9}{8} i \catd \ln^2 2
 - 5 i \cat4 
 + \frac{9}{2} i \Q3 \ln 2
 + 9 i \Q4
   $\\   
$H_{1, 1, -1, 0}(i)  $&$
   \frac{91}{9216} \pi^4 
 - \frac{13}{16} \Z3 \ln 2 
 + \frac{5}{96} \pi^2 \ln^2 2
 - \frac{9}{128} \ln^4 2
 - \frac{27}{16} \A4
 + \frac{1}{4} \catd \pi \ln 2 
 + \frac{1}{2}\pi \Q3
 + \frac{111}{256} i \pi \Z3
 - \frac{1}{192} i \pi^3 \ln 2
   $\\&$
 + \frac{1}{96} i \pi \ln^3 2
 - \frac{1}{8} i \catd \ln^2 2
 - \frac{1}{32} i \catd \pi^2 
 - i \cat4 
 - \frac{1}{2} i \Q3 \ln 2
 - i \Q4
   $\\   
$H_{1, 1, 1, 0}(i)  $&$
   \frac{71}{15360} \pi^4
 - \frac{1}{2} \Z3 \ln 2 
 - \frac{5}{384} \ln^4 2
 - \frac{5}{16} \A4
 + \frac{29}{256} i \pi \Z3 
 + \frac{1}{192} i \pi^3 \ln 2
 - \frac{1}{32} i \catd \pi^2
 - \frac{1}{8} i \catd \ln^2 2
 - \frac{1}{96} i \pi \ln^3 2
   $\\&$
 - \frac{1}{2} i \Q3 \ln 2
 - i \Q4
   $\\   
\hline
$ {\rm Li}_4(1/2)       $&$ a_4$\\
${\rm Im Li}_2 (i)     $&$ \catd$\\
${\rm Im Li}_4 (i)     $&$ \cat4$\\
${\rm Im}H_{0,1,1}(i)   $&$ \Q3$\\
${\rm Im}H_{0,1,1,1}(i) $&$ \Q4$ \\
\hline
\end{tabular}
\end{ruledtabular}
\end{table*}


\end{document}